\documentclass[12pt]{iopart}

\usepackage{graphicx}% Include figure files
\usepackage{dcolumn}% Align table columns on decimal point
\usepackage{bm}% bold math
\usepackage{textcomp}
\usepackage{wasysym}
\usepackage{stmaryrd}
\usepackage{subfig}

\def \ee{\end{equation}}
\def \be{\begin{equation}}

\begin{document}

\title{Exact renormalization group with optimal scale\\
and its application to cosmology}

%\keywords      {Quantum Gravity}
\author{Benjamin Koch and Israel Ramirez}
 \address{
 Pontificia Universidad Cat\'{o}lica de Chile, \\
Av. Vicu\~{n}a Mackenna 4860, \\
Santiago, Chile \\
}
\date{\today}

\begin{abstract}
Assuming an effective gravitational action
with scale dependent coupling constants,
a consistency condition for the local
form of the cut-off scale is derived.
The approach is applied to homogeneous
cosmology and running couplings with an ultraviolet
fixed point. Within the given approach this allows to derive bounds
on the value of the fixed point.
\end{abstract}

\maketitle

%%%%%%%%%% FIGURE %%%%%%%%%%%%%%%%%%

%%%%%%%%%%%%%%%%%%%%%%%%%%%%%%%%%%%%%%%%%%%%

%
%%%%%%%%%%%%%%%%%%%%%%%%%%%%%%%55
\section{Introduction}
Finding a consistent and predictive
description of quatum gravity is a long standing
problem. Aside from the most popular
approaches such as loop quantum gravity, 
the technique of Exact Renormalization Groups (ERG)
has recently attracted more interest.
This technique allows to derive equations
for the running of couplings of an effective average
action without any power expansion in the couplings.
It has been conjectured that the non-renormalizability
of gravity is actually an artefact of the 
loop expansion in the gravitational coupling and
that in an exact calculation the (dimensionless) couplings
actually run to  non trivial ultra violet fixed points.
Since this running would countervail the
divergencies it is called asymptotic 
safety scenario \cite{Weinberg:1979}. In several analytical and numerical
studies supporting evidence for this kind of scenario
has been found 
\cite{Dou:1997fg,Souma:1999at,Reuter:2001ag,Litim:2003vp,Fischer:2006fz,
Litim:2008tt,Narain:2009gb,Groh:2010ta}.
In order to test the implications of this approach 
it has been studied in a large variety of contexts:
Cosmology and Astrophysics 
\cite{Bonanno:2001hi,Bonanno:2001xi,Reuter:2004nx,Reuter:2004nv,Rodrigues:2009vf,Tye:2010an,Bonanno:2010bt},
Brans Dicke theory \cite{Reuter:2003ca},
black holes 
\cite{Bonanno:1998ye,Bonanno:2000ep,Falls:2010he,Reuter:2010xb,Cai:2010zh},
black holes in extra dimensions 
\cite{Koch:2007yt,Koch:2008zzb,Burschil:2009va,Bleicher:2010qr},
interactions in extra dimensions \cite{Fischer:2006fz},
modified dispersion relations \cite{Girelli:2006sc},
f(R) gravity \cite{Codello:2007bd},
deformed special relativity \cite{Calmet:2010tx}, and
gravitational collapse \cite{Casadio:2010fw}.
This work contributes to the above studies in three ways:

 In section \ref{approxSol} a simple expansion 
is used in order to derive a compact
analytic form for the running couplings $G_k$ and $\Lambda_k$
with two fixed point parameters $g^{*}$ and $\lambda^*$. 

 In section \ref{consChap} a new approach for the choice
of the cut-off scale $k$ is proposed.
In the suggested choice, all solutions respect diffeomorphism 
invariance by construction. 
This feature is the main difference from existing studies.

 In section \ref{CosmChap} the findings from the sections
\ref{approxSol} and \ref{consChap}
are applied to cosmology and conditions
on the fixed point parameters $g^*$, $\lambda^*$ are derived and
discussed.

%
%%%%%%%%%%%%%%%%%%%%%%%%%%%%%%%55
\section{An Approximate Analytical Solution to Exact Renormalization Group
Equations}\label{approxSol}

\begin{figure}[hbt]
  \centering
\subfloat[\hspace{7.0cm}.
Analytical solution to the approximated ERGEs for the 
values $g^*=0.016$ and $\lambda^*=0.25$.
Trajectories with $e>0$ imply that $\lambda>0$,
trajectories with $e=0$ imply that $\lambda \rightarrow 0$ 
in the infrared, and
trajectories with $e<0$ give $\lambda>0$ in the ultra violet
and $\lambda<0$ in the infrared.]{\label{gl0}
\includegraphics[width=0.48\textwidth]{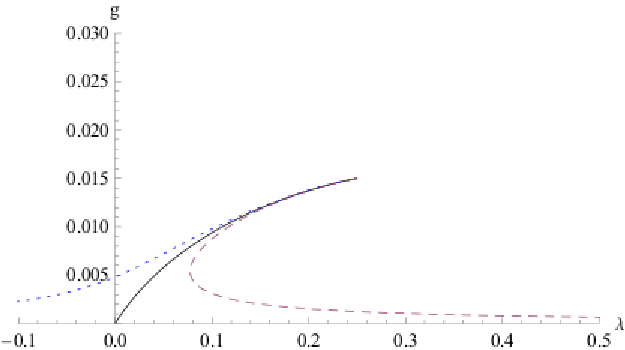}}
\hspace{0.2cm}
\subfloat[\hspace{7cm}.
Numerical solution to the ERGEs.
The choice of the constant $e$ in the
analytical case corresponds to a different choice
of initial conditions in the plane $\lambda$-$g$.]{\label{gl1}
\includegraphics[width=0.48\textwidth]{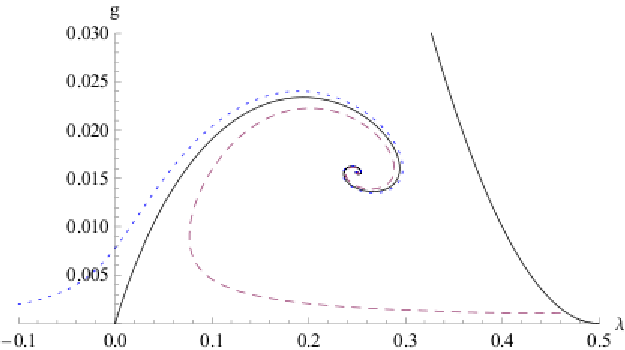}}
\caption{Solutions to the ERGEs (\ref{betal}, \ref{betag}).\label{glboth}
for $G_0=1$.}
\label{gl}
\end{figure}

The Exact Renormalization Group (ERG) was  studied in
the context of effective potentials 
\cite{Wetterich:1992yh}.
This lead to an Exact Renormalization Group Equation (ERGE)
\begin{equation}
 \partial_t\Gamma_k=\frac{1}{2}Tr
\left(\Gamma_{k}^{(2)}+R^{(0)}\right)^{-1}\partial_t
R^{(0)} \label{erge}\quad,
\end{equation}
where $\Gamma_k$ stands for the effective average action, $R^{(0)}$ is the
momentum-cut-off,
and Tr stands for the sum over all indices in DeWitt
notation (see also \cite{Rosten:2010vm}
for a recent review). 
For the case
of quantum gravity in the Einstein-Hilbert tuncation
the effective action in four dimensions is
given by,
\begin{equation}
 \Gamma_k=\int d^4x\frac{\sqrt{g}}{16\pi G_k}\left[R(g)-2
\Lambda_k\right]\quad,
\label{effective}
\end{equation}
where the subscript $k$ denotes that $G$ and $\Lambda$  are scale-dependent
functions \cite{Reuter:2001ag}.
For the ERGE equations (\ref{erge}) and the effective
action (\ref{effective})
one defines 
the dimensionless renormalized gravitational and cosmological constants,
\be\label{Gvong}
 g_k=k^{2} G_k \quad, \quad
\lambda_k=\frac{\Lambda_k}{k^{2}}\quad.
\ee
Using a sharp cut-off \cite{Litim:2003vp}
the beta functions for those constants are found to be
\begin{eqnarray}
 \beta_\lambda&=&\partial_t \lambda_k=\frac{P_1}{P_2+4(4+2g_k)}
 \quad, \label{betal}\\
 \beta_g&=&\partial_t g_k=\frac{2g_kP_2}{P_2+4(4+2g_k)}\label{betag}\quad,
\end{eqnarray}
where $t=\ln k$ and $P_i$ are polynomials of $g_k$ and $\lambda_k$
\cite{Reuter:2001ag,Bonanno:2000ep}.
Those equations have the approximate analytical solution
of the form 
\begin{eqnarray}
g(k)&=&\frac{k^2}{1+k^2/g^*} \quad,\label{gk}\\
 \lambda(g)&=&\frac{g^*\lambda^*}{g}
\left((5+e)\left[1-g/g^*\right]^{3/2}-5+3g/(2g^*)(5-
g/g^*)\right) \quad,\label{lvong}
\end{eqnarray}
which directly defines $\lambda(k)$.
While the functional form of $g(k)$ is frequently used,
$\lambda(g)$ is a new approximation, which is the result
of a Taylor expansion around $\lambda \ll 1$.
The constant $e$ arises as integration constant of the
equations (\ref{betal},\ref{betag}).
As shown in the following study on cosmology
this constant can be fixed from the infrared 
behavior of $\lambda_k$ and
$g_k$.
The constants $\lambda^*$ and $g^*$
are parameterizations of the UV fixed points
\be
g(k^2\rightarrow\infty)=g^*\;\;\mbox{and}\quad
\lambda(k^2\rightarrow\infty)=\lambda^* \quad.
\ee
Numerical results \cite{Fischer:2006fz,Litim:2008tt}, 
reveal evidence for specific values for
the fixed point,
in this work however $\lambda^*$ and $g^*$ will
be treated as free parameters, that 
should be restricted or determined by observational data.
Figure
\ref{glboth} shows the behavior of the numeric and
the analytic solutions (\ref{gk}, \ref{lvong}).
The numerical solution in figure \ref{gl1} is however limited
to the curves on the left of the parabula, where
the beta functions (\ref{betal}, \ref{betag}) become infinite.

Although, those results are obtained in spaces
with Euclidian signature $(+,+,+,+)$, they will also
be used in metric spaces with the signature $(-,+,+,+)$.
%
%
%%%%%%%%%%%%%%%%%%%%%%%%%%%%%%%55
\section{A consistent cut-off scale}
\label{consChap}

Coupling the Einstein-Hilbert action to matter with scale dependent
couplings $\Lambda_k$ and $G_k$ gives 
\begin{eqnarray}
 S[g]&=&\int d^4 x \sqrt{-g}\left(\frac{R-2\Lambda_{ k}}{16\pi G_{
k}}+\mathcal{L}_m\right)\quad. \label{action}
\end{eqnarray}
The equations of motion for the metric field in (\ref{action}) are
\begin{eqnarray}
 G_{\mu\nu}=-g_{\mu\nu}\Lambda_{k}+8\pi G_{k}T_{\mu\nu}-
\Delta t_{\mu \nu}\quad,
\label{eom}
\end{eqnarray}
where the possible coordinate dependence of $G_k$
induces an additional contribution to the stress energy tensor
\cite{Reuter:2004nx,Carroll:2004st}
\be
\Delta t_{\mu \nu}=G_{
k}\left(g_{\mu\nu}\Box-\nabla_\mu\nabla_\nu\right)\frac{1}{G_{k}}\quad.
\ee 
The equations of motion (\ref{eom})
and their solution depend crucially on the 
choice of the cut-off scale $k$.
In particle physics such a scale
would be chosen according to the energy
of the process that is examined. In a curved
spacetime such a definition has no global
unique meaning and the choice of the local
functional form of $k(x^\mu)$ becomes tricky
if not even arbitrary. 
As explained in \cite{Percacci:2007sz} $k(x^\mu)$
has to be guessed for each specific problem.
Typically $k\sim 1/(\int ds)$ is used
for black holes \cite{Bonanno:2000ep,Burschil:2009va}
and  $k\sim 1/t$ or $k\sim 1/H$ is used for
cosmology \cite{Bonanno:2001xi,Bonanno:2010bt}.
Any such choice, however, implies further difficulties:
General covariance dictates that
the equations of motion have to fulfill
the Bianchi identities
\begin{eqnarray}
 G_{\mu\nu}^{\quad;\nu}=g_{\mu_\nu}^{\quad;\nu}=0\quad.
\label{Bianchi}
\end{eqnarray}
We will assume that the conservation of the matter stress energy
tensor persists 
\be\label{Tmunu}
T_{\mu\nu}^{\quad;\nu}=0\quad.
\ee 
Although this assumption is plausible,
it might not be generally true, for example 
it does not hold whenever
particles are created (as it is presumable the case for 
non extremal black holes).
By using (\ref{Tmunu})  and (\ref{Bianchi}) it follows
for the right hand side of (\ref{eom}) that
\begin{eqnarray}
\left(8\pi G_{\bar k}^\prime T_{\mu\nu}-g_{\mu\nu}\Lambda_{\bar
k}^\prime\right)\partial^\nu  k-\nabla^{\nu}\Delta t_{\mu \nu}&=&0
\label{consistency} \quad,
\end{eqnarray}
where
$\partial_\mu G_k=(\partial G_k/\partial k) (\partial_\mu k)=G_k^\prime
\partial_\mu k$ was applied.
In most existing studies
\cite{Bonanno:2001hi,Bonanno:2001xi,Reuter:2004nx,Reuter:2004nx}, 
due the explicit choice of
the functional form of $k(x)$, one always had to
sacrifice either general covariance (\ref{Bianchi}),
or the conservation of the stress energy tensor (\ref{Tmunu}).
In contrast to this, we propose to use the identity (\ref{consistency})
in order to determine the relation between $k$ and $x^\nu$.
Thus, by choosing $k(x)$ to be a solution of the consistency
equation (\ref{consistency}), general covariance and
conservation of the stress-energy tensor can be perserved
by construction.
This procedure is analog to the one that
was discussed in \cite{Babic:2004ev,Domazet:2010bk},
 here however also the 
induced stress-energy tensor $\Delta t_{\mu \nu}$
is taken into account.

In the spirit of \cite{Weinberg:2009wa} it would also be desirable
to have a physical result which does not depend on the local
choice of the cut-off scale $k$.
This can be achieved by treating the function $k$ like a 
classical field and minimizing
the effective action (\ref{action}) with respect to this field.
The corresponding equation of motion for $k$ is
\be\label{eomk}
2\frac{\Lambda_k
G^\prime_k}{G_k^2}-R\frac{G^\prime_k}{G_k^2}-2\frac{\Lambda_k^\prime}{G_k}
=0\quad.
\ee
Although the ``equation of motion'' for $k^2$ (\ref{eomk}) and 
the consistency condition (\ref{consistency})
were derived from completely different motivations,
they are equivalent. This equivalence
can be shown by using (\ref{Bianchi}) and the identity
\be
\nabla^\nu (\nabla_\mu \nabla_\nu-g_{\mu \nu}\Box)\left(
\frac{1}{G_k}\right)=
R_{\nu \mu}\nabla^\nu\left(
\frac{1}{G_k}\right)\quad,
\ee
which allows to rewrite the consistency condition 
(\ref{consistency}) in the form
\be\label{condi222}
R\nabla_\mu \left(\frac{1}{G_k}\right)-
2\nabla_\mu\left(\frac{\Lambda_k}{G_k}\right)=0\quad.
\ee
The above relation is equivalent to (\ref{eomk}),
since the only $x_\mu$
dependence in the couplings comes
due to the functional form of $k^2$.
This can be seen by explicitly 
writing out the covariant
derivatives in (\ref{condi222}).
%The relation also reveals a peculiarity
%of this approach: For models without
%cosmological constant and without any other coupling,
%the gravitational coupling has to be constant.
%
Although a minimal dependence of $k$ and
a consistency of the equations of motion
are desirable,
it is not clear whether they
corresponds to a well defined limit 
in the framework of the exact renormalization group
equation (\ref{erge}). 
This issue arises since
in this limit the effective average action becomes extremal
and the derivation of the ERGE as it is performed in \cite{Reuter:2001ag}
might not be valid any more.
However, in this work it will be supposed
that the solutions (\ref{gk}, \ref{lvong}) are still valid.
%
%%%%%%%%%%%%%%%%%%%%%%%%%%%%%%%%%%%%%%%%
\section{Cosmology}\label{CosmChap}

In this section the concepts that were introduced in the previous sections
will be applied to spatially homogeneous universe,
which can be parameterized by the metric
\begin{eqnarray}\label{metric}
 ds^2=-dt^2+a(t)^2 d\vec x^2 \quad.
\end{eqnarray}
Like in the case of standard cosmology the
matter stress energy tensor will be assumed to
be a perfect fluid.
The homogeneity should also hold for the
running coupling constants and the scale parameter $k=k(t)$.
Having time dependencies only one finds
that the induced stress energy tensor takes the form
\begin{eqnarray}
 \Delta t_{00}&=& -\frac{3 \dot G_k \dot a}{G_k a}\\ \nonumber
 \Delta t_{ii}&=&\frac{a}{G_k^2}(a \ddot G_k G_k-2 \dot G_k^2a+2 \dot G_k \dot a
G_k)\\ \nonumber
\Delta t_{\mu \neq \nu}&=&\quad0 \quad.
\end{eqnarray}
With this the Einstein equations reduce to generalizations of the 
Friedmann equations
\begin{eqnarray}\label{frw1}
 \left(\frac{\dot a}{a} \right)^2 &=&
\frac{8 \pi G_k}{3}\left( \frac{a_0^4 \rho_{r}}{a^4}+
\frac{a_0^3 \rho_{m}}{a^3}\right)+
\frac{\Lambda_k}{3}-\frac{\kappa}{a^2}+
\frac{\dot G_k \dot a}{G_k a} \quad, \\ \label{frw2}
\frac{\ddot a}{a}&=&
-\frac{8\pi G_k}{3}\left( \frac{a_0^4 \rho_{r}}{a^4}+
\frac{a_0^3 \rho_{m}}{2a^3}\right)+
\frac{\Lambda_k}{3}+
\frac{\dot G_k \dot a}{2 G_k a} +
\frac{ G_k \ddot G_k-2\dot G_k^2}{2 G_k^2}\quad. 
\end{eqnarray}
The two equations (\ref{frw1}, \ref{frw2}) would be equivalent
for a fixed coupling $\dot G_k=0$, but they are
not equivalent for the time dependent coupling $\dot G_k\neq 0$.
Without abandoning the conservation of energy
and momentum
this problem can be solved 
if both $G_k$ and $\Lambda_k$
are time dependent quantities \cite{Shapiro:2004ch,Sola:2007sv}.
In the context of ERGE it means that 
the consistency condition (\ref{consistency}) has to be imposed.
For the above equations one finds
\be\label{condiFRW1}
8\pi \dot G_k\left(\frac{a_0^4 \rho_{r}}{a^4}+
\frac{a_0^3 \rho_{m}}{a^3}
\right)+\dot \Lambda_k+
3\dot G_k\frac{\dot a \dot G_k+ G_k \ddot a}{a G_k^2}=0\quad.
\ee
One can show that this condition makes the two equations
(\ref{frw1}, \ref{frw2})
mathematically equivalent and thus it is sufficient just to
work with (\ref{frw1}) and (\ref{condiFRW1}).
By the use of equation (\ref{frw1}), the matter part
in (\ref{condiFRW1}) can be replaced leading to
a more compact form of the consistency condition
\be\label{condiFRW2}
3\alpha(t)-\Lambda_k+G_k \frac{\dot \Lambda_k}{\dot G_k}=0 \quad,
\ee
where we defined $\alpha(t)=(\dot a/a)^2+\ddot a/a+\kappa/a^2$.
Thus, the problem can in principle be reduced to first 
solving (\ref{condiFRW2}) in order to find
the functional relation for $k=k(\alpha)$ and then inserting this
into the first Friedman equation (\ref{frw1}), which then has
to be solved analytically or numerically.

Before starting with this procedure it is interesting
to return for a moment to the equation which was obtained
from demanding a minimal scale dependence of the
effective action. For the homogeneous metric (\ref{metric})
this equation (\ref{eomk}) is exactly the same as
the consistency condition (\ref{condiFRW2}).
This identity is not a coincidence
but rather comes from a deeper connection 
between (\ref{consistency})
and (\ref{eomk}) that was generally shown at the
end of the previous section.

For finding solutions
of the system (\ref{condiFRW2}, \ref{frw1}) an explicit form of
the running couplings $\Lambda_k$ and $G_k$ has 
to be assumed. In order to maintain analytical
feasibility of the problem the approximated solutions
(\ref{gk}, \ref{lvong}) will be used.
The first task is to solve (\ref{condiFRW2}) 
in order to find the functional form
of $k=k(\alpha)$. The solution is however a
relatively large expression, which is not instructive
in this form. Furthermore,  
this algebraic expression has to be inserted
into the generalized Friedman equation
(\ref{frw1}).
This gives a
complicated non-linear differential equation
of higher order, which can probably
not be solved analytically.
Therefore,  the 
model will be studied in the infrared (IR)
and the ultraviolet (UV) limit.
%
%%%%%%%%%%%%%%%%%%%%%%%%%%
\subsection{ERG cosmology in the IR}
In this limit it is assumed that the energy scale
is way below the Planck scale $k^2\ll 1/G_0$.
This allows to expand (\ref{condiFRW2}) in 
a Taylor series around $k^2=0$
\be
3\alpha+\frac{k^2 \lambda^*}{4}(-3+e)+
\frac{e \lambda^* g^*}{2 G_0}=0+{\mathcal{O}}(k^4 G_0^2)\quad.
\ee
which is solved by
$k^2_{IR}=(2e \lambda^*  g^* -12 G_0 \alpha)/((-3+e)G_0 \lambda^*)$.
Since the constants $\lambda^*,g^*,G_0$ are positive one sees
that for $e<3$ the numerator of this solution has to
be less than zero for the square of the energy
scale $k_{IR}^2$ to be positive. For decreasing
$\alpha$ this does not hold any more as soon
as $\alpha \le e \lambda^* g^*/(6 G_0)$.
The solution to this problem is that
from this moment on the IR limit is reached and the
running of the couplings stops.
Thus the infrared solution for $k(\alpha)$ reads
\be\label{k2IR}
k^2_{IR}=\left\{
\begin{array}{cc}
\frac{2 e \lambda^* g^* -12 G_0 \alpha}{(-3+e)G_0 2\lambda^*}
& {\mbox{for}}\;\alpha \ge e \lambda^* g^*/(6 G_0)\\
0& {\mbox{for}}\;\alpha < e \lambda^* g^*/(6 G_0)\quad.
\end{array}
\right.
\ee
Given the typical values for $g^*$ and $\lambda^*$
\cite{Reuter:2001ag,Litim:2003vp} of order one, 
todays value of $\alpha$ definitely demands that $k^2_{today}=0$.
In this limit the modified Friedmann equation (\ref{frw1}) takes the
familiar form
\be\label{frw1IR}
 \left(\frac{\dot a}{a} \right)^2 =
\frac{8 \pi G_0}{3}\left( \frac{a_0^4 \rho_{r}^0}{a^4}+
\frac{a_0^3 \rho_{m}^0}{a^3}\right)+
\frac{e g^* \lambda^*}{3 G_0}-\frac{\kappa}{a^2}\quad.
\ee
This construction is necessary in order
to avoid problems of the big rip type \cite{Caldwell:2003vq}.
For the model to agree with reality,
this equation has to be the same as the standard
Friedmann equation \cite{Friedman:1922a}.
Enforcing this identity one finds that the initially 
arbitrary integration constant $e$ of the exact
renormalization group equations has to be adjusted to the
observed value of the cosmological constant
\be\label{valuee}
e=\Lambda_{observed}\cdot \frac{G_0}{g^* \lambda^*}\quad.
\ee
This is a very strong constraint since it reduces the infinite set
of trajectories that solve (\ref{erge}) to one single curve only.
Within error bars this curve is characterized 
by the positive value given in (\ref{valuee}).
This corresponds to a trajectory
in the figure \ref{gl} where
$\lambda$ is positive for all values of $k$.
As suggested here, the transition between the cosmology
with a variable scale (\ref{frw1}) and
the IR limit (\ref{frw1IR}) is not smooth.
If one sticks to the approximated analytical
form of the couplings as shown in figure \ref{gl0}
it implies violation of total energy and momentum.
This is due to the fact, that the last term
of eq. (\ref{frw1}) still is non zero at the moment
of transition to $k(\alpha_0)=0$.
A straight forward way out of the dilemma is to
conjecture that this additional energy $\delta E $ is transferred to
the matter and radiation content of the universe such that
\be\label{plusDeltas}
\rho_r\rightarrow\rho_r^0=\rho_r+\delta \rho_r\quad
\mbox{and} \quad
\rho_m\rightarrow\rho_m^0=\rho_m+\delta \rho_m \quad.
\ee
Where $\rho_x$ is the density before the transition
and $\rho_x^0$ is the density after
the transition. Such a process can be seen
as the analog of reheating in standard cosmology.
The value of the transition energy will be calculated
in the discussion section.
There is however numerical evidence 
for the curves in figure \ref{gl1} that
$\lim_{\alpha\rightarrow 0}k^2=0$, which
would make the issue of negative
$k^2$ disappear.
%
%%%%%%%%%%%%%%%%%%%%
\subsection{ERG cosmology in the UV}\label{sectUV}
The very early universe was presumably a very
hot environment. Therefore one expects that good estimates
for the Planckian and pre-Planckian epoch can
be obtained by an expansion in ($1/(k^2 G_0)\ll 1$).
In this limit the consistency condition (\ref{condiFRW2})
reads
\be
3\alpha-k^2_{UV}2\lambda^*+
\frac{10\lambda^* g^*}{4 G_0}=0+
{\mathcal{O}}(1/(k^4G_0^2))\quad.
\ee
There is no maximal value for the
scale $k^2$ and thus the asymptotic
solution of (\ref{condiFRW2}) is
\be\label{k2UV}
k^2_{UV}=\frac{3 \alpha}{2\lambda^*}\quad.
\ee
Inserting this solution back into the generalization
of the first Friedman equation
still gives a
non-linear differential equation of third order,
which can not be trivially solved.
However, for asymptotically small times
a solution can be found by making the linear ansatz
\be\label{ansatzUV}
a=C \cdot t \quad,
\ee
where $C$ is a constant that still has to be determined.
With the relation (\ref{k2UV}) and the ansatz
(\ref{ansatzUV}) the differential equation (\ref{frw1})
reduces to
\be
\frac{1}{t^2}=\frac{45 C^4+a_0^4\pi \rho_r \lambda^*32 g^*-9\kappa}
{18 C^2 t^2(C^2+\kappa)}+{\mathcal{O}}(t^0)\quad.
\ee
This proves that (\ref{ansatzUV}) is a solution of the
generalized Friedman equation and that the only
possibly positive linear
expansion coefficient is
\be
C=\frac{1}{3}\sqrt{-3\kappa+2\sqrt{-24a_0^4\pi \rho_r \lambda^* g^*
+9 \kappa^2}}\quad.
\ee
This, implies that this model
only has a physical solution in the UV if
the curvature is different than zero and
fulfills
\begin{eqnarray}\label{limit1}
 \kappa &>& \frac{4}{3}a_0^2 \sqrt{2\pi \rho_r \lambda^*g^*}\quad \quad
\mbox{or}\\ \nonumber
\kappa &<& -2a_0^2 \sqrt{\frac{2}{3}\pi \rho_r \lambda^*g^*}\quad.
\end{eqnarray}
The above inequalities can be translated to dimensionless
matter parameters 
\be
\Omega_r=\frac{8 \pi G_0}{3 H_0^2}a_0^4 \rho_r\;,\quad
\Omega_k=-\frac{\kappa}{H_0^2}\;,\quad
\Omega_\Lambda=\frac{\Lambda}{3 H_0^2}\quad.
\ee
Thus, the relation (\ref{limit1}) restricts the remaining parameters of the running couplings
\be\label{UVineq}
\frac{3}{4}G_0 H_0^2\frac{\Omega_k^2}{\Omega_r}>\lambda^*g^*\quad,
\ee
where it is important to remember, that due
to the energy transfer (\ref{plusDeltas}), the
value of $\Omega_r$ is not necessarily the same
as the one observed today $\Omega_r^0$.

%%%%%%%%%%%%%%%%%%%%%%%%%%%%%%%55
\section{Discussion}
\begin{figure}[hbt]
   \centering
%%\centerline{\protect\vbox{\epsfig{file=fvsr_d2_tt001_Mf1.eps,
%%width=0.6\textwidth}}}
\includegraphics[width=10cm]{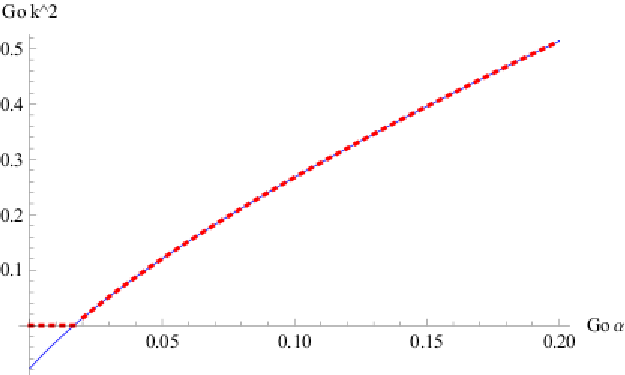}
  \caption{\label{figk22} Functional
dependence of $G_0 k^2$ and $G_0 \alpha$
for $e=0.1$.
The blue line is the complete solution of 
the condition (\ref{condiFRW2}),
while the red-dotted line includes the condition (\ref{k2IR})
in the infrared limit.}
\end{figure}

The behavior of the solution of the condition (\ref{condiFRW2})
can best be demonstrated graphically.
In figure \ref{figk22}, the dependence between
the dimensionless quantities $G_0 k^2$ and $G_0 \alpha$ is shown.
The IR limit corresponds to the left region
of the figure \ref{figk22}.
One sees that it shows no running of the
couplings at the value $k^2=0$, and
a linear dependence between $G_0 k^2$ and $G_0 \alpha$
for small values of $G_0 k^2$.
The UV limit corresponds to the far right hand side of
the plot, with large values of $G_0 k^2$. In this limit
the linear growth continues as it is already visible
on the right hand side of the figure, which reflects
directly the findings of the approximation (\ref{k2UV}).

As explained in section \ref{sectUV},
this UV dependence implies
an epoch of linear expansion in the early universe (\ref{ansatzUV}).
A result which offers a potential
explanation to the homogeneity of the cosmic
microwave background, alternative to inflation.
This is due to the fact that in a linearly
expanding universe the causal horizon scales as
\be
h_c=\int_{t_i}^{t_f}dt \frac{c}{a(t)}=\frac{c}{C}
\left[\ln \left(\frac{t_f}{t_i}\right)\right] \quad.
\ee
In contrast to this, the Hubble Horizon
in this epoch scales as
\be
h_{H}=\frac{1}{t_f-t_i}\int_{t_f}^{t^i}\frac{c}{\dot a}=\frac{c}{C}\quad.
\ee
Therefore, the early epoch of linear expansion
can create arbitrarily high homogeneities for $t_i\rightarrow 0$.

Although this is a nice feature of the model,
the drawbacks become obvious as soon as one
inserts numbers into the relation (\ref{UVineq}).
Using the given limits on
$\Omega_k$ \cite{Vardanyan:2009ft} it 
reveals that the product of the fixed point
parameters $\lambda^*$ and $g^*$ has to be
of the order of $G_0H_0^2\approx 10^{-122}$, 
if the original radiation density
$\Omega_r$ was of the same order of magnitude
as it is today!
When working with the analytical solutions
one has to study whether one
can evade this extreme bound by
assuming that $\Omega_r$ was extremely small,
and all of the radiation density observed today
$\Omega_r^0$ was produced during the transition (\ref{k2IR})
from running couplings to constant couplings at $k^2=0$.
However, estimating the last term in (\ref{frw1})
in a linear expansion solution like (\ref{ansatzUV})
and comparing it at the moment of transition to
the radiation radiation term in (\ref{frw1IR}) in
the same solution gives
\be\label{backdoor1}
\lambda^* g^*\approx \frac{64}{3(3-e)}G_0 H_0^2 \frac{1}{\Omega_r^0}
\quad.
\ee
Thus, evidently, the idea that the observed
radiation density is due to the IR transition
$\Omega_r^0\approx \delta\Omega_r$
also implies values of the parameters 
$g^*\lambda^*\approx 10^{-120}$.
One last backdoor might lie in the possibility
that $e\approx3$ and the numerator of (\ref{backdoor1})
is going to zero.
But remembering the relation (\ref{valuee})
this possibility implies
\be
\lambda^* g^*\approx G_0 H_0^2 \Omega_\Lambda \quad.
\ee
Like before this means that $g^*\lambda^*\approx 10^{-122}$.
Thus, in all possible scenarios
one obtains that the presented model only is in
agreement with observational data in cosmology if
the parameters $g^* \cdot \lambda^*$ are 
extremely small  ($\approx 10^{-122}$).
%%%%%%%%%%%%%%%%%%%%%%%%%%%%%%%55
\section{Conclusion}
A new approach for finding the optimal and
consistent cut off scale $k^2$ for Einstein-Hilbert action is presented.
This is achieved by taking the relation (\ref{eomk})
as algebraic condition for the local functional form of 
the scale $k^2(x)$.
In this context it has been shown that (\ref{eomk}),
which reflects a minimal $k^2$
dependence as suggested by Weinberg \cite{Weinberg:2009wa},
implies the consistency condition (\ref{consistency})
and vice versa.
An elegant way of expressing this condition is given
by eq. (\ref{condi222})
\be\nonumber
R\nabla_\mu \left(\frac{1}{G_k}\right)-
2\nabla_\mu\left(\frac{\Lambda_k}{G_k}\right)=0\quad.
\ee

In order to test the new framework it is applied to cosmology for the case
of the approximate analytic ERG solutions (\ref{gk},\ref{lvong}).
The cosmological model obtained by this procedure
offers some promising properties such as standard Friedmann
cosmology in the IR and a linear expansion in the UV.
However, it is also found that the parameters
of the UV fixed point have to take take extreme values 
$g^*\lambda^*\approx 10^{-120}$ for
the model to be in agreement with observed cosmological
parameters. The order of magnitude of the parameters
reminds of the vacuum density problem in quantum field theory.
Such a small value for the fixed point parameters
does not only appear to be extremely fine tuned, it
is also in disagreement with the findings from approximate 
numerical fixed point searches
\cite{Fischer:2006fz,Litim:2008tt,Codello:2007bd,Groh:2010ta}.
Thus, demanding to have
a classically consistent effective action 
(with demanding (\ref{Tmunu}))
for running dimensionless couplings (\ref{gk}, \ref{lvong}) with a fixed
point at the order of 0.1 does not produce
a good cosmological model.
This allows for at least three different types of conclusions:
\begin{itemize}
\item First, the findings could simply mean that a more complete
cosmological model including other fields (quintessence, inflaton, ...)
is needed. This possibility can be checked by trying more general models,
but it is not very attractive, 
since one has to introduce new parameters which reduces the
predictive power of the theory.
\item Second, the findings could be a hint that there is a generic problem with
the approach. For example it might 
be necessary to relax the assumption of a conserved
stress-energy tensor (\ref{Tmunu})
in the presence of a horizon, where the production
of particles can be expected.
This possibility should be checked by applying the
approach to other problems in general relativity such
as black holes.
\item Third, the findings could be a consequence of the particular
choice that has been made for the functional
form of the running couplings ($G_k$ \& $\Lambda_k$).
In this context, the numerical solution \ref{gl1} should
be studied in comparison with the analytical approximation \ref{gl0}.
Since other candidates for effective theories
of quantum gravity predict different scaling behavior
of the couplings \cite{Nicolini:2008aj,Modesto:2010rv,Garattini:2010dn}, 
they should also be checked directly
within the given approach.
\end{itemize}

Many thanks to M. A. Diaz, A. Gomberoff, and M. Ba\~{n}ados for
helpful hints and discussions.
This work was supported by CONICYT
project PBCTNRO PSD-73 and 
FONDECYT project 1090753.

%%%%%%%%%%%%%%%%%%%%%%%%%%%%%%%%%%%%%%%%%%%55
%\bibliographystyle{h-physrev.bst}
%\bibliographystyle{apsrev}
%\bibliography{../bibfile}
%\begin{thebibliography}{10}

%\end{thebibliography}

\end{document}